# Android Malware Detection Based on RGB Images and Multi-feature Fusion

Zhiqiang Wang *
Department of Cyberspace Security,
Beijing Electronic Science and
Technology Institute
Beijing, China
wangzq@besti.edu.cn

Qiulong Yu *
Department of Cyberspace Security,
Beijing Electronic Science and
Technology Institute
Beijing, China
yuqiulong@163.com

Sicheng Yuan
Department of Cyberspace Security,
Beijing Electronic Science and
Technology Institute
Beijing, China
1343163430@qq.com

*Abstract*—With the widespread adoption of smartphones, Android malware has become a significant challenge in the field of mobile device security. Current Android malware detection methods often rely on feature engineering to construct dynamic or static features, which are then used for learning. However, static feature-based methods struggle to counter code obfuscation, packing, and signing techniques, while dynamic feature-based methods involve time-consuming feature extraction. Image-based methods for Android malware detection offer better resilience against malware variants and polymorphic malware. This paper proposes an end-to-end Android malware detection technique based on RGB images and multi-feature fusion. The approach involves extracting Dalvik Executable (DEX) files, AndroidManifest.xml files, and API calls from APK files, converting them into grayscale images, and enhancing their texture features using Canny edge detection, histogram equalization, and adaptive thresholding techniques. These grayscale images are then combined into an RGB image containing multi-feature fusion information, which is analyzed using mainstream image classification models for Android malware detection. Extensive experiments demonstrate that the proposed method effectively captures Android malware characteristics, achieving an accuracy of up to 97.25%, outperforming existing detection methods that rely solely on DEX files as classification features. Additionally, ablation experiments confirm the effectiveness of using the three key files for feature representation in the proposed approach.

## I. INTRODUCTION

Since its release by Google in 2007, the Android system has quickly dominated the market due to its openness and flexibility. According to data from Stat Counter [1], as of 2024, 71.74% of mobile phones use the Android operating system, and due to its large volume of applications and significant market share, 97% of smartphone malware targets Android devices. This extensive usage and high level of openness have made the Android system a primary target for malware attacks, posing severe security and privacy threats. With the advancement of machine learning and deep learning technologies, using these techniques for malware detection has become a new trend. Existing Android malware detection techniques can be broadly categorized into two types: static analysis and dynamic analysis. However, static analysis struggles against evasion techniques such as code obfuscation, repackaging, and signing, while dynamic analysis is time-consuming and may fail to fully expose some dynamic features. As a result, a novel approach in Android malware detection has emerged in recent years: converting the features of the software under investigation into visual images for analysis and detection. This method not only eliminates the need for expert knowledge but also effectively defends against obfuscation attacks. Meanwhile, Convolutional Neural Networks (CNNs) have demonstrated excellent performance in handling complex image data. Building on this background, this study proposes an Android malware detection method based on RGB images and multi-feature fusion. Compared to traditional methods, our approach converts key file features from APK files into different channels of RGB images and employs feature enhancement techniques, providing a comprehensive and effective representation of Android malware features, thereby improving detection accuracy and robustness.

## II. RESEARCH STATU

In recent years, visualizing Android malware has emerged as an innovative and effective approach. This method leverages the powerful capabilities of computer vision techniques to extract deep features from malware images, enabling more accurate identification and classification of malicious behaviors. Among these techniques, file visualization is a common and effective approach within functional visualization. Unlike other visualization methods, it does not require expert knowledge or manual feature selection (such as extracting variables, function classes, or permissions from Android applications) and can efficiently handle large datasets. Furthermore, different malware variants may share common code modules, leading to similarities in the transformed images, which makes malware visualization particularly useful for classifying Android malware families.

In 2011, Nataraj et al. [2] proposed a novel approach for malware visualization and automatic classification. They treated malware binary files as grayscale images and observed that malware images within the same family exhibit high similarity in layout and texture. Based on this observation, they introduced a classification method using standard image features without the need for code disassembly or execution. This was the first instance of using malware visualization for malware detection.

In 2015, Syed Zainudeen Mohd Shaid et al. [3]introduced a new technique for malware behavior visualization, which displayed the behavioral characteristics of malware through behavior images for identifying malware variants. Their approach first captured malware behavior by executing it in a virtual machine and then transformed these behaviors into images through color mapping.

---

* Co-first Author.

In 2018, TonTon Hsien-De Huang and Hung-Yu Kao [4] proposed a color-inspired Convolutional Neural Network (CNN) method for detecting Android malware. This method, named R2-D2, converts the bytecode of Android application Class.dex files into RGB color codes, generating fixed-size color images. These color images are then used as input for the CNN, enabling automatic feature extraction and training.

In 2020, Jinrong Chen et al. [5] proposed a method that converts malware opcodes and data flows into RGB pixels, generating RGB images, which are then used to recognize and classify malware using Convolutional Neural Networks (CNNs) from deep learning. Additionally, Abir Rahali et al. [6] introduced a novel Android malware detection system named DIDroid, which utilizes deep image learning techniques with a large labeled dataset comprising 200,000 malicious and 200,000 benign samples. This approach demonstrated the effectiveness of deep image learning in classifying and characterizing Android malware. By converting APK file features into two-dimensional images and employing CNNs, the method achieved a detection accuracy of 93.36% with training and test log losses below 0.20.

In 2021, Nadia Daoudi et al. [7] presented a new Android malware detection method based on deep learning. Their method, named DexRay, converts the bytecode of Android application DEX files into grayscale "vector" images, which are then processed using a one-dimensional Convolutional Neural Network model. The simplicity of this method is its innovation, as it directly utilizes raw bytecode without the need for complex feature engineering.

In 2022, Suleiman Y. Yerima and Abul Bashar [8] combined image-based features with features from the Manifest file, employing Histogram of Oriented Gradients (HOG) and byte histograms to extract features from images representing application executable files. These features were then combined with features from the Manifest file for classification using machine learning algorithms.

In 2023, Zhao et al. [9] transformed malware samples into RGB images and applied transfer learning on ResNet networks to train and classify these RGB images.

In 2024, Amel Ksibi et al. [10] utilized pre-trained deep learning models such as DenseNet169, Xception, InceptionV3, ResNet50, and VGG16 for feature extraction. They aimed to train deep learning models by converting Android APK files into binary code and RGB images. These models were trained and evaluated on the CICInvesAndMal2019 dataset, demonstrating an accuracy of up to 95.83% in distinguishing between malicious and benign applications.

III. RELEVANT KNOWLEDGE

Before you begin to format your paper, first write and save the content as a separate text file. Complete all content and organizational editing before formatting. Please note sections A-D below for more information on proofreading, spelling and grammar.

The Android operating system is structured in a layered architecture. From the bottom to the top, it primarily includes the following layers: the Kernel Layer, Hardware Abstraction Layer (HAL), System Libraries and Android Runtime Environment, Application Framework Layer, and Application Layer. As shown in Fig. 1 below.

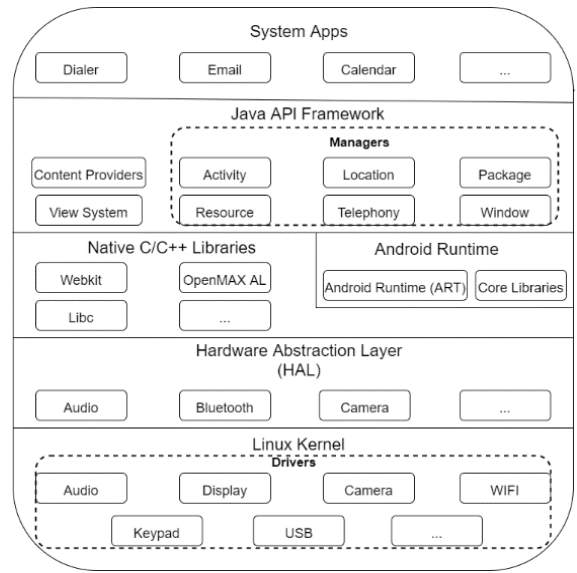

Fig. 1. Android Layered Architecture Diagram

The Kernel Layer is based on the Linux kernel, which is responsible for driving the underlying hardware and ensuring the basic operation of the system, providing core functionalities. The Hardware Abstraction Layer offers a clear interface that is independent of hardware for the upper Android system, allowing system services and the application framework to operate on different hardware platforms. The System Libraries, consisting of C/C++ libraries, support the operation of system components, while the Android Runtime Environment provides the Android Runtime (ART), which implements core functionalities of the Java language. The Application Framework Layer includes various system-level services and management components, such as the Window Manager and Notification Manager, offering a rich set of API interfaces that enable developers to create feature-rich applications. The topmost Application Layer is the user-facing interface, encompassing all user and system applications.

A. *Relevant Component Features*

   *1) DEX Files*

   In the Android system, a DEX (Dalvik Executable) file is a specialized file format used to store compiled code for Android applications. DEX files are designed for the Dalvik Virtual Machine, which was a unique runtime environment in earlier versions of Android but has since been replaced by the Android Runtime (ART). Despite this replacement, the DEX file format continues to be used. Within an application's APK file, the DEX file contains all the executable code, including class definitions, method implementations, and other code structure information.

   *2) AndroidManifest.xml File*

   The AndroidManifest.xml file is located in the root directory of an application project and serves as the global configuration file for the application. It defines the structure and metadata of the application and informs the Android operating system on how to execute the application. The AndroidManifest.xml file includes information on application permissions, component declarations, application requirements, and features. It is designed to centrally manage critical information about the application, ensuring that the Android system can effectively parse and run the application. Given that it contains detailed structural information and permission requests about the application, it is often used as a key resource for security analysis.

*3) API Calls*

In programming and application development, API (Application Programming Interface) calls are a crucial mechanism that allows interaction between software components. In the Android system, API calls are particularly vital as they enable applications to access the core functionalities and services provided by the Android platform, such as accessing device hardware, managing system resources, and handling network communications. As shown in Fig. 2 below.

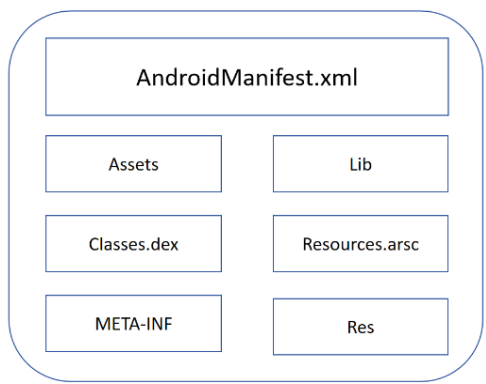

Fig. 2. APK File Structure Diagram

B. *Image Enhancement Techniques*

*1) Canny Edge Detection*

Canny edge detection [11] is a widely used image processing technique, proposed by John Canny in 1986. It aims to efficiently identify edges in images and is a fundamental technique in the field of image processing, particularly important in visual object recognition and scene analysis. The primary advantage of the Canny edge detection algorithm is its ability to provide good noise suppression while maintaining high precision in edge localization. The process of Canny edge detection mainly involves steps such as noise reduction, gradient magnitude and direction calculation, non-maximum suppression, double threshold detection, edge linking, and edge tracking.

By applying Canny edge detection to process DEX files, it can highlight key features of the code structure, thereby assisting in analyzing software behavioral characteristics and improving the accuracy of malware detection.

*2) Histogram Equalization*

Histogram equalization is a commonly used technique in image processing to improve the contrast of an image. This technique adjusts the intensity distribution of an image, making the intensity more evenly distributed across the image, thus enhancing the visual effect. Histogram equalization is particularly suitable for images where both the background and the foreground are too dark or too bright. It automatically adjusts the contrast of the image, making the details more clearly visible. The basic steps include calculating the histogram, computing the cumulative distribution function (CDF), mapping new pixel values, and creating a new image.

Histogram equalization can better showcase details in both bright and dark areas of an image, making it useful in applications such as medical imaging, satellite image processing, and any scenario where improved image quality is needed for more precise analysis. In Android application analysis, applying histogram equalization to the AndroidManifest.xml file can more clearly reveal the structure and details within the XML file, aiding in the identification and analysis of critical information.

*3) Adaptive Thresholding*

Adaptive thresholding is an image processing technique used to convert an image into a binary image, where each pixel has only two possible intensity values (typically black and white). Unlike simple global thresholding, adaptive thresholding determines the threshold for each pixel based on the pixel values in its surrounding area, making the algorithm more flexible and effective when dealing with local variations in lighting conditions within the image. The basic steps include selecting the window size, calculating the local threshold, and applying the threshold.

Applying adaptive thresholding to the grayscale images generated from API calls can better handle variations caused by complex backgrounds or uneven lighting conditions, making important API call patterns more prominent. This technique helps to more accurately identify critical regions within the image, thereby enhancing the effectiveness of malware detection algorithms.

C. *Deep Learning Models*

In this study, we employed several advanced deep learning models to validate the effectiveness of the multi-feature fusion method. These models include AlexNet, GoogleNet, ResNet, ResMLP, and MobileNetV2.

AlexNet [12] is a milestone in deep convolutional neural networks, achieving a breakthrough performance in the 2012 ImageNet Challenge, significantly improving the accuracy of image classification tasks. The model consists of five convolutional layers and three fully connected layers, utilizing ReLU as the nonlinear activation function, effectively avoiding the vanishing gradient problem. AlexNet also introduced local response normalization (LRN) and overlapping max pooling to enhance the model's generalization capability. Additionally, it used dropout techniques to reduce overfitting, which was an important innovation in neural network training at the time.

GoogleNet [13] introduced an innovative module called Inception, which allows the network to automatically learn the most suitable combination of convolution kernel sizes. Each Inception module contains multiple parallel convolutional layers and pooling layers, enabling the capture of information at different scales within the same layer. GoogleNet has a total of 22 layers, deeper than previous networks, yet with fewer parameters, thanks to the clever design of the Inception module, which significantly reduces the computational resources required.

ResNet [14] addresses the degradation problem in deep neural networks by introducing a residual learning framework. The core idea is to use residual modules, allowing the input to be directly passed to the subsequent layers through cross-layer connections. This design enables the network to learn the residual mapping between the input and output, simplifying the learning objective and optimization process. ResNet can build very deep networks (e.g., ResNet-152) and has achieved outstanding performance in several critical visual recognition tasks.

ResMLP[15] is an architecture based on multilayer perceptrons (MLPs), replacing traditional convolution operations with fully connected layers in each block. By introducing residual connections and LayerNorm, ResMLP can effectively perform deep feature learning. This architecture demonstrates the potential of fully connected networks in handling visual tasks, providing an efficient alternative, especially in resource-constrained scenarios.

MobileNetV2[16] is a lightweight deep convolutional neural network designed for efficient computation on mobile and embedded devices. Proposed in 2018, the model introduces a technique called depthwise separable convolution, significantly reducing the number of parameters and computational cost while maintaining high classification accuracy. MobileNetV2 employs an inverted residual structure and linear bottleneck layers to further improve computational efficiency. The inverted residual structure reduces computational complexity by performing convolution operations in a low-dimensional space, while the linear bottleneck layer avoids information loss. This design allows MobileNetV2 to achieve efficient inference on resource-constrained devices, making it widely applicable in various scenarios with limited computational resources, such as mobile devices and IoT devices.

IV. DESIGN AND IMPLEMENTATION OF THE METHOD

This paper proposes an Android malware detection method based on RGB images and multi-feature fusion. The method involves extracting three key types of data from APK files—DEX files, AndroidManifest.xml files, and API calls—converting them into grayscale images, enhancing their features, and then combining them into an RGB image. This RGB image is subsequently used with mainstream image classification models for malware detection and classification. Specifically, Canny edge detection is applied to the grayscale image of the DEX files to enhance edge features; histogram equalization is used on the grayscale image of the AndroidManifest.xml files to improve contrast; and adaptive thresholding is applied to the grayscale image of API calls to highlight important call patterns. These processed grayscale images are then merged into an RGB image, with each channel representing one of the feature types. This method integrates various image processing techniques to effectively enhance feature recognition, thereby improving the accuracy and robustness of malware detection and classification. The architecture diagram of the proposed malware detection scheme is shown in Fig 3.

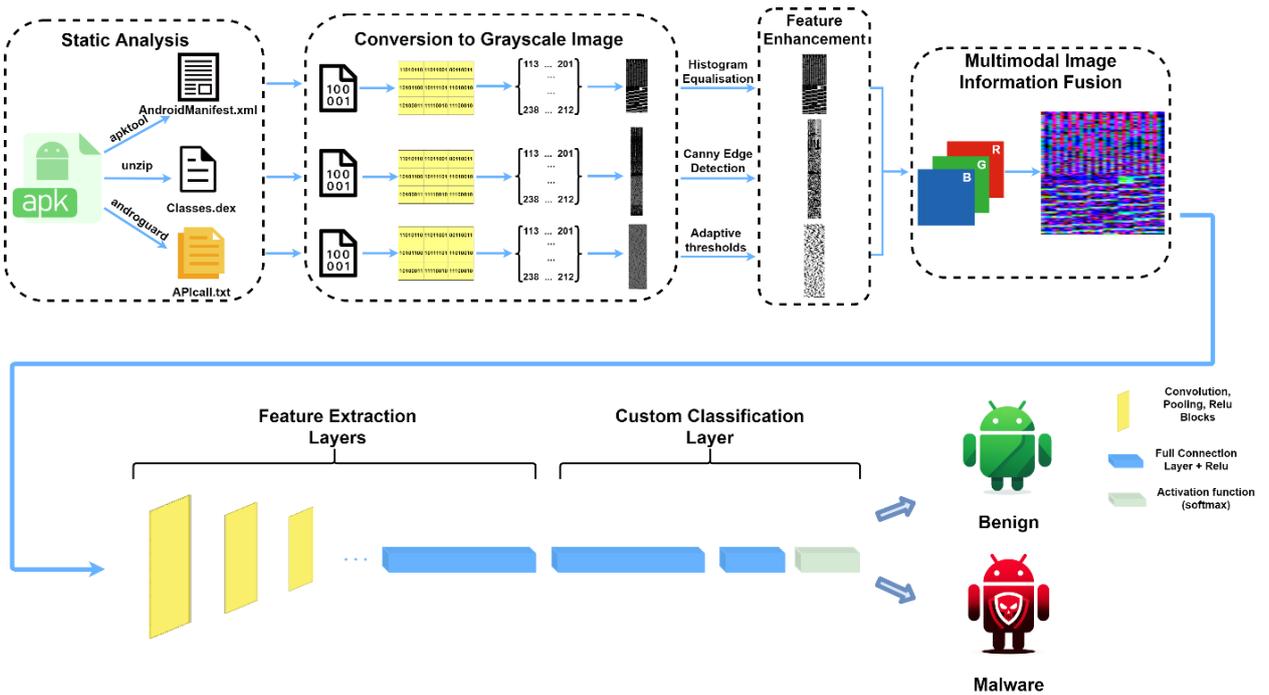

Fig. 3. Proposed Scheme Architecture Diagram

A. File Extraction

 1) Extraction of DEX Files

Firstly, the classes.dex file is extracted from the APK. An APK is essentially a ZIP archive, and in this study, the ZipFile's read() method is used to directly read and extract the classes.dex file from the APK. The ZipFile is used to iterate over the APK files, reading the contents of each file sequentially. If a file with the .dex extension is encountered, the read() function is used to read the content, which is then saved as an external file with the same name as the original APK.

 2) Extraction of AndroidManifest.xml

The AndroidManifest.xml file is extracted using the Apktool tool, which can completely extract files such as resources and AndroidManifest.xml from the APK package. The d (decode) command of Apktool is used to unpack the target APK file. This process extracts the APK file into a specified directory and converts all resources, code, and metadata within the package into an editable and viewable format. After unpacking, Apktool automatically converts the AndroidManifest.xml file from its original binary XML format to a more readable and editable plain XML format. The converted file retains all application declarations, including permission requirements, defined components, and activities.

*3) Extraction of API Calls*

API call information is extracted from the APK file using the Androguard tool. Androguard is designed for in-depth analysis of Android applications and can retrieve detailed information about API calls, including system-level calls such as network access and system service utilization, as well as third-party library calls. The decompiled information is organized into a text file.

*B. Conversion of File Information to Grayscale Images*

Grayscale images are a special format where each pixel value represents only grayscale information, encompassing shades of gray from black to white, without color information. In a grayscale image, pixel values typically range from 0 to 255, where 0 represents pure black, 255 represents pure white, and intermediate values represent varying shades of gray. Converting binary data to grayscale images essentially maps numerical information to a visual representation.

The extracted classes.dex, AndroidManifest.xml, and API call information from the APK exist in binary form, containing critical details such as application execution logic, permission requirements, and interface call patterns.

The extracted binary data is then converted to grayscale values according to predefined mapping rules. In this study, each byte of binary data is interpreted as an integer between 0 and 255, directly corresponding to the pixel values in the grayscale image. This method ensures the directness and completeness of data conversion, with each byte's value accurately represented in the converted image.

Determining the dimensions of the converted image is a crucial step. Considering the size differences among various files, a dynamic calculation method is employed to determine the width and height of the image, ensuring that all extracted data is appropriately mapped to the grayscale image while maintaining the aspect ratio as close as possible to the original data's structural features.

Based on the determined image dimensions and mapping rules, each extracted information item (classes.dex, AndroidManifest.xml, and API calls) is converted into independent grayscale images.

Converting binary data to grayscale images allows the abstract numerical information to be observed and analyzed in an intuitive form and facilitates the application of deep learning models for malware detection. Given the strong performance of deep learning techniques in image processing and recognition, these technologies can be used to identify complex patterns and features from the converted grayscale images, thus effectively detecting and classifying Android malware.

*C. Grayscale Image Feature Enhancement*

For the grayscale images converted from the classes.dex files, the Canny edge detection algorithm is applied to prominently identify the edges of the code structure. Each image file is read in grayscale mode, and the Canny function uses two threshold parameters (set to 100 and 200 in this study) to determine which regions of the image have sufficiently significant pixel intensity changes to be marked as edges.

For the grayscale images converted from AndroidManifest.xml, histogram equalization is applied to enhance the image contrast. This processing aims to improve the visual information in the image, particularly the clarity of structure and text. The cv2.equalizeHist function is used for histogram equalization, which computes the image histogram and redistributes pixel intensities to enhance overall contrast. This process makes details that may have been obscured by uneven lighting or other factors more visible, particularly the text and label structures in AndroidManifest.xml, which is crucial for subsequent text parsing and feature extraction.

Additionally, for the grayscale images converted from API calls, adaptive thresholding is used to better highlight the API call patterns in the image. This method is particularly suitable for images with uneven lighting or background changes, as it adjusts the threshold based on local image characteristics, thus preserving image details while reducing noise. The cv2.adaptiveThreshold function is applied, with the maximum threshold set to 255, the adaptive method using a Gaussian window (cv2.ADAPTIVE_THRESH_GAUSSIAN_C), the threshold type set to binary threshold (cv2.THRESH_BINARY), the neighborhood size set to 11, and the constant C set to 2, which is subtracted from the computed mean or weighted mean to determine the final threshold. By adjusting these parameters, the threshold for each pixel can be dynamically tuned to better highlight the API call patterns.

*D. Multi-feature Fusion*

In this study, we developed an automated process to convert multi-feature fusion extracted from Android application packages (APKs)—specifically DEX files, AndroidManifest.xml configuration files, and API calls—into RGB images for deep learning model analysis. This process involves initially converting these different forms of data into grayscale images and then merging them into a three-channel RGB image, with each channel representing a different feature visually.

The implementation process begins with identifying and reading the three types of feature files from a specified directory: PNG images of the DEX files, PNG images of the AndroidManifest.xml file, and PNG images of the API calls. These files are named to reflect the APK name and its classification label, ensuring accurate association in subsequent processing steps. The PNG images are opened in grayscale mode ("L" mode) using the Image module from the Pillow library (PIL).

To address potential size mismatches after converting different source files into images, we use a scaling approach. The resize_image function is employed to adjust the images to a predefined target size (256x256 pixels, determined to yield optimal detection results through comparative experiments). In this process, we use the Lanczos resampling algorithm, a high-quality resampling filter suited for image scaling, which preserves the original image details while minimizing distortion introduced during resizing.

After resizing each image to the same dimensions, we use the Image.merge method from Pillow to combine the three grayscale images into a single RGB image. In this merging process, the grayscale image of the DEX file is placed in the red channel, the grayscale image of the AndroidManifest.xml

file is placed in the green channel, and the grayscale image of the API calls is placed in the blue channel. By assigning each feature to a specific color channel in the final RGB image, this approach captures and represents the multi-feature fusion information in an integrated visual format.

*E. Malware Detection*

To demonstrate the effectiveness and generalizability of the proposed multi-feature fusion information fusion approach, various baseline models with different architectures were employed for Android malware detection, including AlexNet, GoogleNet, ResNet, MobileNetV2, and ResMLP. MobileNetV2, AlexNet, GoogleNet, and ResNet are CNN-based models. After fusing multi-feature fusion information into RGB images, these models extract local features through convolutional layers, perform downsampling with pooling layers, and finally classify the images using fully connected layers. MobileNetV2 incorporates depthwise separable convolutions to reduce computational complexity, GoogleNet introduces multi-scale feature extraction, and ResNet employs residual connections to address the challenges of training deep networks. ResMLP, on the other hand, is based on a multilayer perceptron (MLP) architecture, utilizing fully connected layers and residual connections for feature extraction.

## V. EXPERIMENTAL RESULTS AND COMPARISON

*A. Dataset*

This study utilized the CICMalDroid 2020 dataset[17] for model training and evaluation. CICMalDroid 2020 is a comprehensive Android malware dataset released by the Canadian Institute for Cybersecurity (CIC). The dataset encompasses a wide range of Android malware samples, covering various types of malicious behaviors.

TABLE I. DATASET DISTRIBUTION

| Sample Type | Quantity |
|---|---|
| Benign Samples | 4039 |
| Malware Samples | 12623 |

The Android application dataset includes a total of over 16,000 applications across five major categories: 4039 benign applications, 1512 adware, 2467 banking malware, 3896 mobile riskware, and 4809 SMS malware.

*B. Experimental Environment and Evaluation Metrics*

The experiments were conducted using an Ubuntu 22.04 system with the PyTorch 1.11 deep learning framework. GPU acceleration with CUDA 11.5 was employed to speed up neural network training, and all parts of the experiment were implemented in Python.

When using deep learning models for Android malware detection, four types of outcomes can occur: True Negative (TN), True Positive (TP), False Negative (FN), and False Positive (FP). Specifically, TN refers to benign samples predicted as benign, TP refers to malicious samples predicted as malicious, FN refers to malicious samples predicted as benign, and FP refers to benign samples predicted as malicious. The evaluation metrics used for the model in this study are as follows: Accuracy, Precision, Recall, and F1 Score, calculated as follows:

$$\text{Accuracy} = \frac{TP + TN}{TP + TN + FP + FN} \quad (1)$$

$$\text{Precision} = \frac{TP}{TP + FP} \quad (2)$$

$$\text{Recall} = \frac{TP}{TP + FN} \quad (3)$$

$$F_1\text{-score} = 2 \times \frac{\text{Precision} \times \text{Recall}}{\text{Precision} + \text{Recall}} \quad (4)$$

*C. Comparison with Feature Extraction Based Solely on DEX Files*

Since DEX files integrate all class definitions, method codes, fields, and other information of an application, using DEX files for feature extraction and transformation is one of the most commonly employed methods in image-based Android malware detection, as demonstrated in references [4] and [7].

Under the condition that the CICMalDroid 2020 dataset, model architecture, and training hyperparameters are consistently set, we compared our feature extraction and transformation method with those from references [4] and [7]. The experimental results are presented in Table II and Figure 4.

TABLE II. COMPARISON WITH FEATURE EXTRACTION BASED SOLELY ON DEX FILES

| | Accuracy | Precision | Recall | F1-score |
|---|---|---|---|---|
| Alexnet-Literature [4] | 96.014% | 92.53% | 95.59% | 93.95% |
| Alexnet-Literature [7] | 93.564% | 92.75% | 85.93% | 88.82% |
| Alexnet-Our method | 96.700% | 94.10% | 95.59% | 94.82% |
| Googlenet-Literature [4] | 93.499% | 88.69% | 91.18% | 89.86% |
| Googlenet-Literature [7] | 91.408% | 84.81% | 91.19% | 87.40% |
| Googlene-Our method | 95.198% | 92.10% | 92.27% | 92.19% |
| Resnet-Literature [4] | 94.381% | 88.91% | 95.08% | 91.52% |
| Resnet-Literature [7] | 96.635% | 95.35% | 93.76% | 94.53% |
| Resnet-Our method | 96.896% | 95.50% | 94.21% | 94.84% |
| Resmlp-Literature [4] | 93.662% | 92.89% | 85.81% | 88.80% |
| Resmlp-Literature [7] | 94.414% | 90.77% | 91.74% | 91.24% |
| Resmlp-Our method | 95.557% | 92.67% | 93.96% | 93.30% |
| Mobilenetv2-Literature [4] | 96.960% | 95.53% | 95.04% | 95.28% |
| Mobilenetv2-Literature [7] | 96.831% | 95.06% | 94.65% | 94.85% |
| Mobilenetv2-Our method | 97.131% | 96.20% | 94.87% | 95.52% |

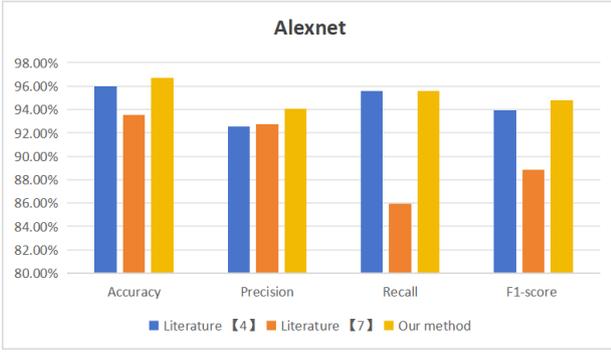

Fig. 4(a). Comparison of Alexnet experiments

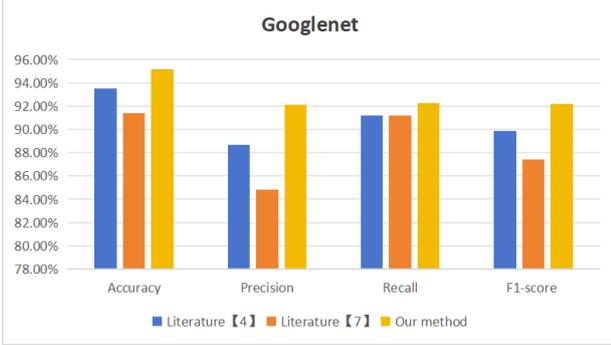

Fig. 4(b). Comparison of Googlenet experiments

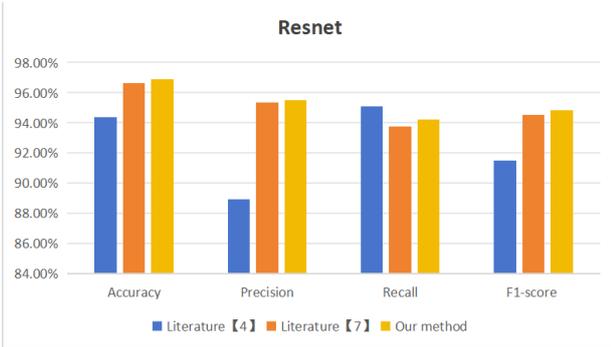

Fig. 4(c). Comparison of Resnet experiments

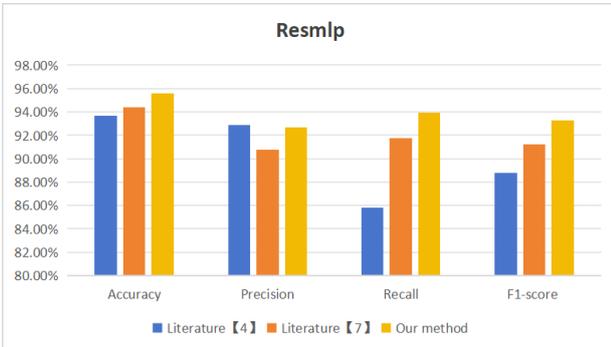

Fig. 4(d). Comparison of Resmlp experiments

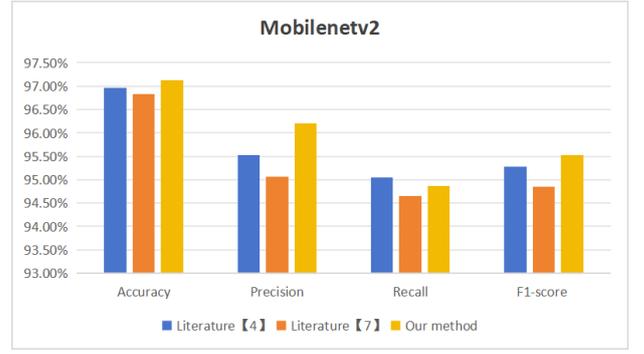

Fig. 4(e). Comparison of Mobilenetv2 experiments

The experiments indicate that RGB images, which integrate multi-feature fusion information, provide better predictive performance compared to predictions based solely on features extracted from DEX files. Our method demonstrates superior performance across various test models, showing significant improvements particularly in Accuracy, Precision, and F1-Score.

This suggests that the fusion of features extracted from DEX files, AndroidManifest.xml files, and API calls enables a more comprehensive capture of APK file characteristics, thereby enhancing the accuracy and reliability of malware detection.

*D. Ablation Experiment*

To validate the effectiveness of the proposed method, we conducted ablation experiments to assess the contribution of each individual image channel to malware detection performance. Specifically, we removed each of the following channels—DEX file channel, AndroidManifest.xml file channel, and API call file channel—and performed experiments using the remaining two types of files for feature extraction. The feature sets are defined as follows:

TABLE III. DEFINITION OF FEATURE SETS

| Feature Set | Feature Type |
|---|---|
| Feature Set I | DEX File |
| Feature Set II | Manifest.xml File |
| Feature Set III | API Call File |

The results of the ablation experiments using GoogLeNet and ResNet models are shown in the following tables and Figure 5:

TABLE IV. ABLATION EXPERIMENT RESULTS WITH GOOGLENET

| Feature Set Combination | accuracy | precision | recall | f1-score |
|---|---|---|---|---|
| I+II | 94.022% | 92.56% | 87.9% | 89.99% |
| I+III | 95.328% | 92.17% | 92.7% | 92.47% |
| II+III | 93.924% | 89.91% | 91.6% | 90.73% |
| I+II+III | **95.655%** | **93.02%** | **92.9%** | **92.99%** |

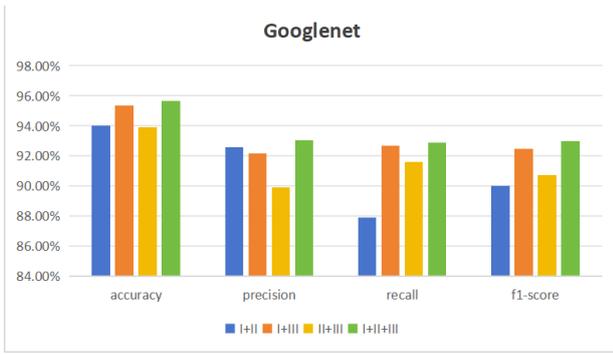

Fig. 5(a). Ablation Experiment Results with GoogLeNet

TABLE V. ABLATION EXPERIMENT RESULTS WITH RESNET

| Feature Set Combination | accuracy | precision | recall | f1-score |
|---|---|---|---|---|
| I+II | 97.190% | 96.14% | 94.84% | 95.47% |
| I+III | 96.210% | 95.95% | 92.22% | 93.93% |
| II+III | 96.014% | 92.84% | 94.77% | 93.76% |
| I+II+III | **97.256%** | 95.66% | **95.77%** | **95.71%** |

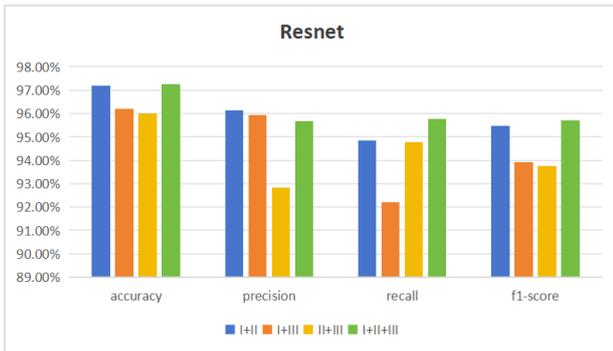

Fig. 5(b). Ablation Experiment Results with ResNet

The results demonstrate that multi-feature fusion RGB images generated from the fusion of DEX files, AndroidManifest.xml, and API calls outperform the use of any single channel in malware detection. The performance of the model decreases when any one channel is removed, highlighting the significant contribution of each data type to enhancing the model's predictive capability. Specifically, the performance decline is most pronounced when the DEX channel is removed, indicating that the execution code information in DEX files is crucial for the prediction model. Similarly, the configuration information from AndroidManifest.xml and the behavioral data from API calls underscore their essential role in providing contextual and behavioral features. Thus, these experimental results confirm that integrating multiple data channels can substantially improve the accuracy and robustness of malware detection. This finding provides valuable insights for designing more effective Android malware detection solutions in complex data environments.

## VI. CONCLUSION

In this study, we explored an Android malware detection method based on RGB images and multi-feature fusion. By extracting DEX files, AndroidManifest.xml files, and API calls from APK files, we converted each into grayscale images, applied specific image enhancement techniques to improve feature representation, and then fused them into RGB images for classification and detection. The experimental results demonstrate that our method significantly outperforms the commonly used approach of converting only classes.dex into RGB images, showing notable improvements in detection performance across models such as AlexNet, GoogLeNet, MobileNetV2, ResNet, and ResMLP. Additionally, ablation experiments further validated the contribution of each channel to detection performance, revealing that removing any grayscale image channel results in decreased detection accuracy. This indicates that the multi-feature fusion approach is more effective. Future work will focus on further optimizing the malware detection models to enhance accuracy, exploring additional effective feature extraction methods from malware, and improving data processing workflows to achieve more efficient malware detection.